\documentclass[]{aastex631}

\shorttitle{Accretion in a triple system as an origin of luminous hyper-soft sources}
\shortauthors{S.B. Popov and G.V. Lipunova}
\graphicspath{{./}{figures/}}

\begin{document}

\title{Disc accretion onto a binary black hole in a hierarchical triple system as an origin of the most luminous hyper-soft sources}

\correspondingauthor{Sergei Popov}
\email{sergepolar@gmail.com}

\author[0000-0002-4292-8638]{Sergei B. Popov}
\affiliation{Sternberg Astronomical Institute, Universitetski 13, Moscow, 119234, Russia}

\author[0000-0003-4515-8955]{Galina V. Lipunova}
\affiliation{Max-Planck-Institut für Radioastronomie, Auf dem Hügel 69, Bonn, 53121, Germany}



\begin{abstract}

 We propose that the recently discovered luminous hypersoft X-ray sources can be explained by accretion onto a binary black hole in a hierarchical
 triple system. For black hole masses  $\sim 15 M_\odot$, the orbital separation of the internal binary might be $\sim 0.01 $~AU. If the donor provides $\gtrsim 10^{-8} M_\odot$ yr$^{-1}$, then the circumbinary accretion 
 disc can explain the observed properties of the most luminous supersoft sources.

\end{abstract}




\section{Introduction} 
\label{sec:intro}

{Recently,} 
\cite{2026arXiv260206192M}
presented a population of hypersoft X-ray sources (HSS). They are extragalactic non-nuclear objects selected on the basis of non-detection above 0.3~keV in Chandra observations
(photon ratio 
for the energy ranges 0.15-0.3 keV and 0.3-1 keV is $\gtrsim$ 8).  
The whole sample contains 84 sources in a wide range of soft X-ray luminosities. Here, we are mostly
interested in sources with the largest 
bolometric 
luminosities, as their origin remains uncertain.

Some sources in the presented sample have estimated bolometric luminosities $\gtrsim10^{39}$~erg~s$^{-1}$. 
In the case of accretion, 
such luminosities are super-Eddington for accreting white dwarfs and neutron stars. 
Thus, 
black holes (BHs) might be considered as accretors.
Furthermore, 
observations point to
thermal spectra with $T_\mathrm{eff}\approx 10$~--~20~eV. 
This allows estimating the size of the emitting area: $R_\mathrm{em}\sim 10^{11}-10^{12}$~cm.  
Inner parts of accretion discs around stellar mass BHs are expected to be 
sufficiently
smaller and hotter. 
The probability of having an intermediate mass BH as an accretor in a binary system is very small \citep{2007MNRAS.377..835K}, in contradiction to the large number of HSS detected per galaxy. 
To explain the main properties of high-luminosity HSS, 
 we propose a model of accretion onto a binary BH in a hierarchical triple.\footnote{This is an expanded version, {dated 27.02.2026}, of a brief contribution published in Res. Notes AAS, 10, 46 (2026).} 

\section{Parameters of the disc around a binary black hole}

We consider a stellar interacting hierarchical triple system,
where
mass transfer proceeds from an optical 
a donor star
via Roche lobe overflow through the inner Lagrange point 
or via intense stellar wind.
{However, instead of accreting onto a single compact object, the transferred material feeds a binary BH.}
The gas forms an emitting 
a circumbinary accretion
disc surrounding the BH pair.
The semimajor axis of the BH binary, $a_0$, can be restricted from above 
by the size of the emitting region: $\lesssim 10^{12}$~cm. 
A circular-orbit binary coalesces due to gravitational waves (GWs) emission 
in:
\begin{equation}\label{eq.tau_coal}
    \tau_\mathrm{coal}=\frac{5}{256} \frac{a_0^4}{G^3} \frac{c^5}{M_1 M_2(M_1+M_2)}
    \approx 1.6\times 10^{6} \, {\mathrm{yrs}} \, \left(\frac{a_0}{0.01\, \mathrm{AU}}\right)^4, 
\end{equation}
for $M_1=M_2=10\, M_\odot$. 
Thus, 
we expect $a_0\gtrsim 0.01$~AU.
Thus, stellar-mass BH pairs with separations less than $0.01$ AU are effectively unobservable due to a very short lifetime as $\tau_\mathrm{coal}\propto a^4$.
%
{However (see below), in some cases, the inner binary separation can decrease or increase due to the interaction with the circumbinary disc. Correspondingly, this would modify the lifetime of the system.} 


The orbital evolution of the binary is influenced by GWs and angular momentum exchange with the circumbinary disc via resonant gravitational torques and accretion into a cavity. 
For certain combinations of binary parameters (total mass, mass ratio, and separation) and disc properties, the inflow of matter from the circumbinary disc onto the individual BHs can be strongly suppressed.
{As angular momentum is transferred from the binary to the disc, the binary orbit contracts.
In the opposite case, when significant accretion of mass and angular momentum occurs across the cavity onto the two BHs, the total angular momentum of the binary can increase. As a consequence, the orbital separation grows rather than shrinks, which would increase the lifetime of the binary.}
In this case, the lower bound on $a_0$ might be lower than provided in the previous paragraph.

It was suggested that a critical $H/R$ value dividing the binary evolution between expansion and contraction lies around $H/R=0.04-0.2$~\citep{Heath-Nixon2020}, where $H (R)$ is the disc semi-thickness, and $R$ is the distance from the center of mass.  For thinner circumbinary discs, accretion across the central cavity is negligible.  This would naturally explain the absence of significant X-ray emission from mini-discs around the BHs in HSS in our framework. The observed 
luminosity is {instead} 
produced by viscous dissipation within the circumbinary disc itself.

Following {the arguments of} 
\citet{Heath-Nixon2020}, we assume that the disc aspect ratio does not exceed a critical value 
0.2 and that its inner radius satisfies $ R_{\mathrm{in}} \gtrsim 2\,a_0$.
Under these conditions, 
we derive constraints on the disc mass and radial extent, as well as on the total mass of two BHs. {These constraints, in turn, allow us to infer plausible properties of the third component of the system --- the donor star responsible for supplying mass to the circumbinary disc.}

The spectrum peaks at $\approx 10$~eV, corresponding to an effective temperature $T_{\rm eff} \approx 10^5$~K. In the estimates below, we assume this value as fiducial. 
From the observed luminosity and $T_\mathrm{eff}$, related by $L \sim \pi R_{\rm em}^2 \, \sigma \, T_{\rm eff}^4$, we find  
the characteristic radius 
$R_{\rm em} \sim 2.4 \times 10^{11} L_{39}^{1/2} T_{\rm eff, 5}^{-2}$~cm (we follow the convention $V_\mathrm{x}=V/10^x$). Accordingly, 
we introduce the scale of the problem, $R_{11} = R/10^{11}~$cm.

The luminosity of the viscous Keplerian disc with the inner radius $R_{\rm in}$ 
around a binary BH with a negligible accretion rate onto the center can be approximated as 
\begin{equation}\label{eq.L_Fvis}
    L = \left(F_{\rm vis, in} \, + \frac{\dot M\, \sqrt{G\,M\,R_{\rm in}}}{2} \right)\omega_{\rm K}(R_{\rm in})\, .
\end{equation}
Here, $\omega_{\rm K}(R_{\rm in})$ and $F_{\rm vis, in}$ are the Keplerian angular velocity  and the viscous torque at $R_{\rm in}$, respectively. 

The viscous torque is defined from the viscous stress tensor as $F_{\rm vis} \equiv 2\,\pi\, W_{r\varphi}\, R^2$~\citep[e.g.,][]{2018ASSL..454....1L}. At the inner edge of a disc around a single BH, which coincides with the innermost stable circular orbit, 
 the stress-tensor is zero, and so is the viscous torque. The inner torque in a disc around a binary BH is provided by the resonant torque. In a stationary viscous disc, the angular momentum conservation yields the distribution of the torque over the radius:
$F_{\rm vis}(r) = F_{\rm vis, in} + \dot M (\sqrt{GMR}-\sqrt{GMR_{\rm in}})$. Evidently, the viscous torque is almost constant if the accretion rate $\dot M $ is small enough.
Such {discs} 
were proposed by \citet{1977PAZh....3..262S} 
{ and called} {\it disc reservoirs}.  {In such viscous reservoirs, the effective temperature varies with radius as $T_{\rm eff} \propto R^{-7/8}$.}

 Equation~(\ref{eq.L_Fvis}) allows us to 
 express the viscous torque from the observables (luminosity and temperature), 
 With an assumption about the magnitude of the viscosity, we 
 obtain the surface density and thickness of the disc.
In $\alpha$-discs, the surface density $\Sigma$  is related to the torque as at each radius as 
\begin{equation}\label{eq.Fvis}
    F_{\rm vis}  = 3\pi\,\alpha\, G\, M\, R\, \left(\frac{H}{R}\right)^2\, \Sigma \, .
 \end{equation}
Actually, in disc reservoirs 
the surface density varies as 
$\Sigma \propto R^{-1} $ in the case of constant aspect ratio $H/R$. Eq.~(\ref{eq.Fvis}) allows us to express   the surface density at the inner edge of the disc (assuming $\dot M=0$ in Eq.~(\ref{eq.L_Fvis})) from the observed 
luminosity $L$ :
  \begin{equation}\label{eq.Sigma}
 \Sigma (R_{\rm in})  \approx 9.4 \times 10^{4}\, {\rm g~cm}^{-2}\, L_{39}^{5/4}\, m_{20}^{-3/2} \, T_5^{-1}\,\alpha_{-1}^{-1}\, \left(\frac{H/R}{0.2}\right) ^{-2}.
 \end{equation}
 Here $M=M_1+M_2= M_1(1+q)$ is the total mass of BHs 
 and $m_{20}=M/20\,M_\odot$.

The cooling of the disc is sustained locally by the radiation diffusion.   The temperature in the disc symmetry plane $T_{\rm c}$ and the effective temperature 
$T_{\rm eff}=T_5 \,10^5$~K are thus related~
\citep[][equation 1.99]{2018ASSL..454....1L}:
\begin{equation}\label{eq.Tc}
    T_{\rm c}^4 \approx \frac 15\, T_{\rm eff}^4 \,\Sigma\, \kappa_{\rm T}\, .
 \end{equation}
 Here  
 $\kappa_{\rm T}\approx 0.35$~cm$^2$ g$^{-1}$ is the scattering cross-section for the solar chemical abundances ($\mu =0.64$).
 We assume that the scattering opacity dominates.

Using a hydrostatic balance, we establish a connection between the sound velocity in the disc plane and the disc thickness. This relation allows us to effectively close the system of equations:
\begin{equation}\label{eq.H_R}
  \frac{H}{R} \approx 2.5  \frac{\sqrt{\Re T_{\rm c}/\mu}}{\omega_{\rm K} R}  .
 \end{equation}
 Here, $\Re$ is the universal gas constant.
 
 Substituting (\ref{eq.Sigma}) into (\ref{eq.Tc}) and subsequently into (\ref{eq.H_R}), we {can} 
 estimate the total mass of the two BHs:
\begin{equation}
     m_{20}  \approx 1.7 \, T_{5}^{-10/11} \,L_{39}^{13/22} \,\kappa_{\rm T}^{2/11}\,  \alpha_{-1}^{-2/11}\, \left(\frac{H/R}{0.2}\right) ^{-20/11}\, .
 \end{equation}

 For a disc with $\dot M = 0$ and with the constant aspect ratio $H/R$, one easily finds 
 $M_{\rm disc} \approx \pi\, \Sigma (R_{\rm in})\, R_{\rm in}\, R_{\rm out} $ or
 \begin{equation}
     M_{\rm disc} \sim 8 \times 10^{-6} \, M_{\odot} \,\frac{L_{39}^{9/4}\, } { T_5^5\, m_{20}^{3/2}\,  \alpha_{-1}\,} \left(\frac{H/R}{0.2}\right) ^{-2} \,\frac{R_{\rm out}}{R_{\rm in}}\, .
 \end{equation}

{Resonant torques that truncate the disc from inside transfer binary angular momentum into the disc and, thus,} 
the binary's orbit shrinks 
on the characteristic 
timescale:
 \begin{equation}
   t_{\rm tid,shrink}  
   =  \frac{\rm {Binary~angular~momentum}}{\rm {Flux~of~angular~momentum~to~disc}} 
   = \frac{M \,  \sqrt{G M a_0}\, q}{F_{\rm vis,in} (1+q)^2} \approx 1.4 \times 10^4\, \mathrm{yr}\, \frac{m_{20}^2\, q \, \sqrt{a_0/R_{\rm in}}}{(1+q)^2  \, L_{39}^{3/2}}.
   \label{tid}
 \end{equation}
 Here, we {took} that 
 the angular momentum of a binary due to its orbital motion is  $M q\sqrt{GM a_0}/(1+q)^2$. 
{Comparison of (\ref{tid}) and {GW coalescence time} 
{(\ref{eq.tau_coal}) 
shows that the GW radiation}
is a minor channel for shrinking a binary at the considered stage.}

  In a low-mass X-ray 
  binary, a mass transfer rate of $\sim 10^{-9}-10^{-8}\,M_{\odot}\,\mathrm{yr}^{-1}$ would need to operate for $\sim 10^4-10^{5}\,\mathrm{yr}$ in order 
  to accumulate the required mass of the disc with $R_{\rm out}/R_{\rm in} \sim 10$. 
  This corresponds to a critical aspect ratio 
  $H/R \sim 0.2$.  \citep{Heath-Nixon2020}, 
 above which mass leakage into the central cavity begins, and a quasi-stationary state may be established.
The viscous time of the disc reservoir, which is about $ M_{\rm disc}\sqrt{G M R_{\rm out}}/F_{\rm vis}$~\citep{2015ApJ...804...87L}, is comparable to (\ref{eq.tau_coal}), ensuring that the BH binary is not decoupled from the disc \citep{2005ApJ...622L..93M}. 
Still, limitations from Eq.~(\ref{tid}) can provide additional constraints.  

{The final state of the evolution of such a triple system involves a BH coalescence with a probable $\gamma$-ray burst (depending on the existence of the mini-discs) and a subsequent X-ray outburst, similar to a scenario proposed for supermassive BHs \citep{2005ApJ...622L..93M}.}


\subsection{ Accretion into a cavity  needed to prevent coalescence}

Accretion into the cavity brings in the angular momentum  $\dot M \sqrt{GM R_{\rm in}}$, which can be transferred to the binary orbital motion in the case of the prompt sink of the gas onto BHs~\citep{2017MNRAS.469.4258T}.  Therefore, to avoid coalescence  {due to GW radiation}, the characteristic time of the angular momentum gain should be comparable to the time (\ref{eq.tau_coal}):
\begin{equation}
\frac{M \,  \sqrt{a_0}\, q}{\dot M \sqrt{R_{\rm in}} (1+q)^2} \lesssim \tau_{\rm coal}
\end{equation}
or
$\dot M \gtrsim 2\times 10^{-6}\, m_{20}^4 \,M_\odot/$yr for $q=1$, $R_{\rm in} = 2\, a_0$. 

However, the resonant torque interaction of the BH pair with the massive circumbinary disc significantly shortens the lifetime of the inner binary, {for the required luminosity, see Eq.~(\ref{tid}).} 
{The accretion rate into the cavity necessary to counterbalance} this shrinking is very high:
$\sim F_{\rm vis,in}/\sqrt{GMR_{\rm in}} \sim 1.4\times 10^{-3} L_{39}^{3/2}\,T_5^{-2}\, m_{20}^{-1}~M_\odot\, {\rm yr}^{-1}$. 
{Apparently, } accretion onto the BHs from the circumbinary disc cannot make the lifetime of the pair much longer than provided by Eq.~(\ref{tid}).



\section{Conclusions} 

Despite some fine-tuning being necessary, an HSS system can represent a triple with accretion onto a central binary BH. The BHs' masses are about $15 M_\odot$ and the orbital separation about $10^{11}$~cm. In $10^4-10^5$~yrs an accretion disc with the necessary properties can be formed from a donor with the mass-capture rate $\gtrsim 10^{-9}-10^{-8} M_\odot$/yr. 
The lifetime of such systems is short 
and the necessary combination of parameters can be quite rare, but calculations of the birthrate of these triples are beyond the scope of this research note.

\bibliography{rnaas_hss}{}
\bibliographystyle{aasjournal}

\end{document}